# Radio-Frequency Rectification on Membrane Bound Pores

Sujatha Ramachandran, Robert H. Blick[*], and Daniel W. van der Weide

*Electrical & Computer Engineering, University of Wisconsin-Madison, Madison, WI 53706-1691, USA.*

We present measurements on direct radio-frequency pumping of ion channels and pores bound in bilipid membranes. We make use of newly developed microcoaxes, which allow delivering the high frequency signal in close proximity to the membrane bound proteins and ion channels. We find rectification of the radio-frequency signal, which is used to pump ions through the channels and pores.

(*) – address for correspondence: blick@engr.wisc.edu

Probing the interaction of biological systems with radio frequencies holds great promise for research and drug screening applications. While a common assumption is that biological systems do not operate at radio frequencies, we find that currents due to ion transport through channels and pores in cell membranes are in the pA to nA range. These values translate via the average current $<I> = <n> e / \tau_d = <n> e f$ to frequencies in the range of 1 MHz to 1 GHz, where $<n>$ is the average number of ions transported per interval and $\tau_d$ is the dwell time of the ions in the channel. It is thus desirable to have circuitry available, which facilitates radio frequency spectroscopy of ion transport with the potential to deliver real-time *in-vitro* information on ion channel operation. Here we present measurements on the local interaction of a radio frequency signal with single ion channels and pores. We find radio frequency rectification and pumping on the channels and pores embedded in suspended bilipid membranes, recorded in direct current measurements. This electromagnetic modulation can be used to probe the dynamics of ion channel conformational changes.

In studying the interaction of electromagnetic signals with biological systems the first consideration is to compare the Debye screening length of the molecule or protein under investigation in aqueous solution to the frequency of operation. For radio frequency (RF) signals this results in very strong attenuation due to the large dielectric constant of water ($\varepsilon \sim 80$). This in turn requires higher signal power, heating the biological system. Hence, heating and rupture of membranes was found in previous work studying the effects of radio and microwave frequencies on ion channels in membranes [1,2,3]. Another hindrance is the low temporal resolution of probing single pore signals with bandwidths of $\sim 10$ kHz. This is due to large access resistances ($R$) and membrane capacitance ($C$), leading to low time constants $\tau = RC$. We overcome these limitations by using microcoaxial lines for delivering RF signals directly to ion channels and pore proteins embedded in suspended bilipid membranes, resulting in pumping of ions through the channels.

The measurement setup is depicted in Fig. 1(a): a chamber combines a standard planar bilayer recording setup with a delrin cuvette for ion transport measurements on bilipid membranes with microcoaxial lines. The chamber has *cis*-and *trans*-side corresponding to drain and source contacts containing 1 M KCl and 10 mM HEPES (pH 7.3) in aqueous solution. The delrin cuvette has a $\sim 200$ micron pore onto which a lipid bilayer (1,2-diphytanoyl phosphatidyl choline in decane) is painted. These membranes are ideal model systems, since they only contain phospholipids and no other proteins, which make them more stable for voltage-current recordings over a long period of time. The DC contacts (Ag/AgCl-wires) are simply dipped into the solution (*cis* side grounded), similar to a standard membrane patch measurement [4]. A patch clamp amplifier Axopatch 200 B (Axon Instruments, Inc., Union City, CA) was used for all the direct current measurements shown. The currents were low-pass filtered with a built-in four pole Bessel filter at 1 kHz and sampled at 10 kHz by computer with a Digidata 1320 analog-to-digital converter. The AC contacts (RF microcoaxes) are brought in from the sides on optical rails to allow for exact positioning close to the suspended bilipid membrane. The alignment procedure is critical for this measurement, since it will determine the coupling efficiency of the RF coaxial probes. The RF generator used was a Hewlett-Packard synthesizer (HP 83650A) generating a sinusoidal continuous wave excitation.

The whole chamber is mounted in a Faraday cage to achieve optimal shielding from spurious electromagnetic radiation. In Fig. 1(b) a magnified image of a typical microcoax is depicted: the outer conductor is separated by a standard dielectric from the inner conductor, which tapers down to a tip of radius just below 1 micron. The whole coaxial tip, which is immersed into the ionic solution of the *cis*-side chamber, is covered by parylene to avoid surface potentials interfering with the DC measurement. Finally, the coaxial tip is mounted and positioned close to the bilipid membrane, about 10 – 15 µm away, as sketched in Fig. 1(c). The thicker edges and thin center of the membrane indicate the typical thinning process of painted membranes. In Fig. 1(d) an equivalent circuit is given: as seen, the two AC lines are DC-blocked allowing the measurement of direct RF transmission. However, in the data discussed here we focus on AC excitation and DC-detection, i.e. only one microcoax is brought to action.

We first defined a stable bilayer and then took current recordings of the insertion of the channel-forming proteins. Only then the RF signal was switched on at a specific frequency and power level. We investigated two types of membrane proteins: Alamethicin (ALA) a fungal peptide which forms voltage-gated ion channels [5] and α-Hemolysin (α-HL) which is a pore forming agent [6]. At a current level for ALA and α-HL of 20 – 50 pA the ion transfer times correspond to values around (1/100) – (1/300) MHz$^{-1}$.

For calibration purposes we first painted a bilipid membrane across the 200 micron aperture and recorded the direct current vs. time trace shown in Fig. 2(a). The insulation resistance of the pure membrane is of the order of 16 GΩ (a bias of –40 mV is applied). Switching the RF source on we find a strong crosstalk signal, which decays with a typical time constant of about one second. With a sinusoidal RF signal at 800 MHz the membrane resistance is reduced to about 8 GΩ, as seen on the right-hand-side of Fig. 2(a). The time constant corresponds to the decharging time of the membrane, taking the membrane resistance of now $R_m$ = 8 GΩ in the on-state and a calculated membrane capacitance of $C_m$ = 137 pF, we find a time constant of $\tau_m = R_m C_m \sim 1$ s. For calculations we assumed the membrane dielectric constant to be $\varepsilon_m \sim 2$ and a circular geometry of the membrane layer with diameter 200 micron. After this initial spike the current relaxes to a non-zero value, already indicating an effective bias across the membrane. Although the power of the RF signal is comparatively high with about $P = -15$ dBm, we find only minimal heating effects as evident in the increase of the DC-current signal noise, indicated by the pairs of arrows left and right, as seen in the broadening of the current vs. time trace.

The initial measurements were performed on ALA : these peptides were added to the *cis*-chamber and can be inserted into the membrane by a small DC-bias voltage, due to their dipole moment of ~75 Dy. In the membrane ALAs can combine in different numbers and form ion channels, as sketched in Fig. 2(b). Depending on the number of peptides current through the ion channel scales accordingly to the cross sectional surface. These channels are only temporarily stable, i.e. their formation depends on the Brownian motion of the peptides in the membrane. The resulting current under a bias voltage of –20 mV is shown in Fig. 2(c). On the left side the typical ALA-recording is found with predominantly smaller channels forming. After switching the RF signal on at 800 MHz and relaxation of the current spike we find a strongly enhanced current level. This already is a clear indication of an effective increase in DC-bias by the voltage. This voltage enhances the current through the membrane by pumping the ions more effectively and increasing channel formation. We found several frequencies (between 750-850 MHz) at which the current through the pores could be increased, but best coupling was achieved at a frequency of 800 MHz. All measurements were executed at this frequency. This is due to the geometry of the conducting elements and the dielectrics of the measurement cell, i.e. the location of the DC lines, the coaxes, the chip, the pore and membrane and the aqueous solution in the chamber. Importantly, the RF signal at 800 MHz is faster and able to reverse (or pump) the flow of ions through the channels and pores with dwell times of $1/100 – 1/200$ MHz$^{-1}$.

In Fig. 3 the measurements are repeated for different DC-bias values: in (a) the conditions are identical except that the bias is inverted to + 20 mV. The ALA peptides are inserted into the membrane and show the typical current spikes. However, when the RF signal is turned on, ion channel formation is knocked off. Only after the excitation is turned off, the ALA channels form again with the typical current trace. In Fig. 3(b) the measurements are repeated with zero bias applied. At zero bias no ALAs are inserted into the membrane and consequently no current spikes are obtained. Switching the RF signal on induces ion channel formation with an average channel current corresponding to an effective bias of about – 25 mV. We can conclude that the application of an appropriate RF modulation corresponds to an effective DC-voltage, which is applied to the membrane. This can be interpreted as rectification of the AC signal at the highly resistive bilipid membrane junction.

Conventionally, the term rectification refers to the conversion of an electromagnetic AC signal to a DC voltage across a junction. The junction's *IV*-characteristic has to be nonlinear of some form, which is the result of asymmetric barrier resistances and capacitances such as in a Schottky-diode. In DC measurements on ALA it was shown by Woolley *et al*. [7] that the *IV*-relation of ALA is non-ohmic due to the electrostatic potential profile within this ion channel – this is also termed rectification. In more recent work by Siwy *et al.* [8,9,10] asymmetric artificial nanopores were fabricated in thin membranes. A similar rectification of ionic currents was found, being related to the asymmetry of the surface charges inside the nanopore. Heins *et al.* [11] finally have shown that this rectification effect can be enhanced by placing asymmetric molecular groups at the nanopore. This so-called chemical rectification leads to a further increase of the non-ohmic response.

Obviously, the asymmetric potential along the ion channel or nanopore determines the transport properties, i.e. the rectification effects should be reflected in the response to an AC excitation. This appears to be the case for the ion channel ALA. In order to determine how the potential and geometrical shape of the protein forming channel influences this rectification process and the resulting pumping of ions, other proteins have to be considered. Hence, we also employed the porin α-HL for the second line of experiments, which in contrast to ALA acts more like a DC resistor once inserted into the bilipid membrane [13,14]. This is related to the fact that pores allow a constant flux of ions after insertion. There is no gating mechanism as for the artificial ion channel ALA. Consequently, pores or *porins,* as they are sometimes called, lend themselves as a perfect 'calibration set'. Furthermore, it is important to note that α-HL is a protein which possesses a highly asymmetric molecular structure, once inserted into the membrane (see Fig. 4(a)). The opening of the pore on the *cis* side has a diameter of 2.9 nm which widens into a mushroom-shaped vestibule of 4.2 nm in diameter and again narrows to an opening of 2 nm on the *trans* side.

In Fig. 4(b) a simplified circuit diagram is shown: in parallel to the membrane resistance and capacitance we place a single α-HL pore with a resistance of $R_{\alpha\text{-HL}}$ = 1.1 GΩ. Once α-HL is added to the *cis*-compartment of the setup it is inserted into the membrane with the head of the protein facing the RF signal and the current level rises to –35 pA under a bias of –40 mV. In Fig. 4(c) the direct current through the pore is shown with and without the RF signal. Evidently, switching the RF on results in a current spike, but with reduced relaxation time as compared to the ALA measurements (Figs. 2,3). This is related to the initial resistance of the system being an order of magnitude smaller $R_{\alpha\text{-HL}} \sim 1$ G$\Omega <$ 9 G$\Omega \sim R_{ALA}$ and consequently $\tau_m \sim$ 100 ms. During RF emission the current through the pore is enhanced by 20 pA, indicating pumping of ions through the pore at a rate of $\tau = f^{-1} = 1/(800$ MHz$)$. This shows the possibility of real-time spectroscopy, i.e. the passage of single ions can now be resolved in the time domain. The current level of |20 pA| then translates to $n = I/(ef) = 0.16$ ions on average, which are pumped through the pore in each cycle of the RF signal ($f = 800$ MHz). The increase by almost a factor of two in current again shows pumping of ions by the RF voltage. It is accompanied by a moderate increase in noise level.

So far the origin of the observed rectification process has not been sufficiently clarified. Considering in a simplified model the flux $J_f$ of ions through an ion channel or pore it should follow in general the Nernst-Planck equation

$$J_f = -\left(\frac{z\Gamma\Delta\varphi}{\Re T}\right)\left(\frac{c_{cis}\exp\left(\frac{z\Gamma\Delta\varphi}{\Re T}\right) - c_{trans}}{\exp\left(\frac{z\Gamma\Delta\varphi}{\Re T}\right) - 1}\right), \qquad (1)$$

where $c_{trans}$ and $c_{cis}$ are the ionic solution concentrations in the cis- and the trans-side, $z$ is the valence of the ion, $\Gamma$ Faraday's constant, $\Re$ the gas constant, and $T$ the temperature. The potential gradient $\Delta\phi$ applied along the ion channel or pore in this approximation is assumed to be linear. Obviously, Eq. (1) is a nonlinear relation for the ionic flux based on the different salt concentrations in both reservoirs. The relation of ion flux to the direct current is given by $I = z\Gamma J_f$, where we quoted only one ion concentration for simplicity. However, in the measurements conducted here we adjusted the ion concentrations so that $c_{trans} = c_{cis}$. This leaves a linear relation for the channel current

$$I = z\Gamma J_f = -\left(\left[\frac{z^2 \Gamma^2}{\Re T}\right]\Delta\phi\right) = R^{-1} V, \qquad (2)$$

with the resistance $R$ and assuming a linear potential drop $\Delta\phi(x) = V$, we find Ohm's law. Now rectification can only occur, if the potential along the channel $\phi(x)$ leads to a nonlinear potential drop, as is the case for ALA [7,12] or if the nonlinearity is caused by an asymmetry of the molecular groups of the pore, as is the case for α-HL [13]. Hence, we can conclude that RF rectification occurs due to the combination of a channel's potential or shape asymmetry and the high overall junction resistance. In this case an effective DC voltage arises and ions are actively pumped through the channel.

In summary we have shown that coated microcoaxes can be applied successfully for delivering RF signals to ion channels and pores in bilipid membranes. We find rectification of the RF voltage with a resulting voltage, which is used for pumping ions through the channels. This method is shown to have potential for real-time spectroscopy on molecular channels and for drug screening applications.

*Acknowledgements* – We like to thank Hagan Bayley for providing the α-HL pores and Meyer Jackson for helpful discussions. This work has been supported in part by the Defense Advanced Research Projects Agency under the MOLDICE grant. DWvdW thanks NIH for an STTR grant in collaboration with Prairie Technologies.

*Figure captions –*

**Figure 1.** Experimental setup: (a) measurement chamber within a Faraday-cage. Two coaxial lines enter from the top, feeding the microcoaxes in the chamber (RF in/out). The direct current through the membrane is probed via Ag/AgCl-wires dipped into the *cis-* and *trans*-chambers. (b) Close-up of the microcoax tip showing electrodes, insulating layer, and outer conductor. The whole microcoax is coated with an insulator (parylene). (c) Configuration in the experiment where the microcoax is placed as close as possible at the suspended membrane, which is sketched spanning the aperture. (d) Circuit diagram for the AC/DC combination.

**Figure 2.** (a) Calibration measurement with a suspended bilipid membrane painted over a 200 micron opening. Shown is the current vs. time recording before and after the radio frequency (RF) signal at 800 MHz is turned on. This indicates a very good electrical insulation of *cis-* and *trans*-side and minimal heating. (b) Sketch of six alamethicin peptides inserted into the suspended lipid bilayer. These peptides conjugate and form channels of different sizes. (c) Recording under –20 mV DC bias and 800 MHz applied. RF is rectified and enhances the formation and current level of alamethicin channels, i.e. ions are more effectively pumped through the membrane.

**Figure 3.** Bias variation on alamethicin under excitation: (a) At +20 mV bias the RF knocks off channel activity, leading to an effective pumping of ions against the bias. (b) RF triggers channel insertion at zero bias applied, i.e. rectification. Both results indicate rectification of the RF signal, since an effective DC voltage is present.

**Figure 4.** Measurements on a protein pore: (a) Single α-HL after insertion into the bilipid membrane from the *cis*-side. The mushroom-like head of the protein faces the incoming RF signal. (b) Circuit diagram with membrane resistance and capacitance, $R_m$ and $C_m$ respectively, and the α-HL as a lumped resistor with $R_{\alpha\text{-HL}} \sim 1.1$ GΩ. (c) Current through a single α-HL pore with the standard level at –35 pA, which is enhanced to about –55 pA under RF pumping at –40 mV bias.

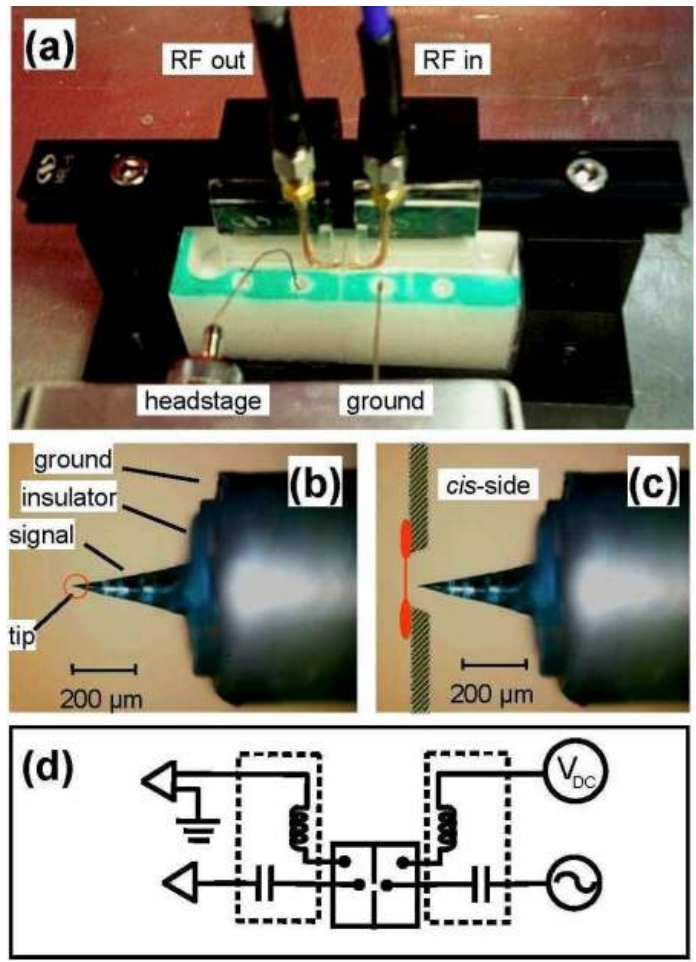

Ramachandran *et al*, Fig. 1/4

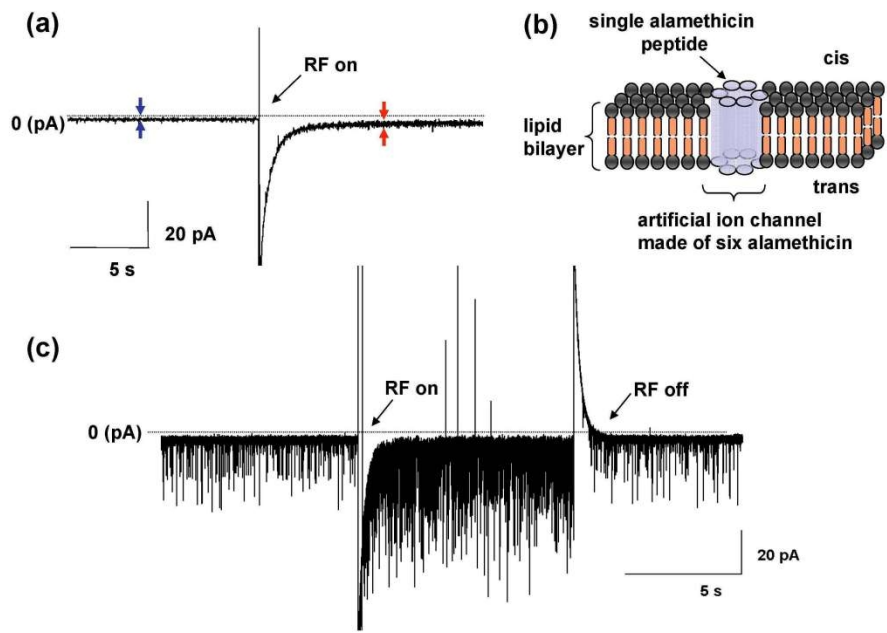

Ramachandran *et al*, Fig. 2/4

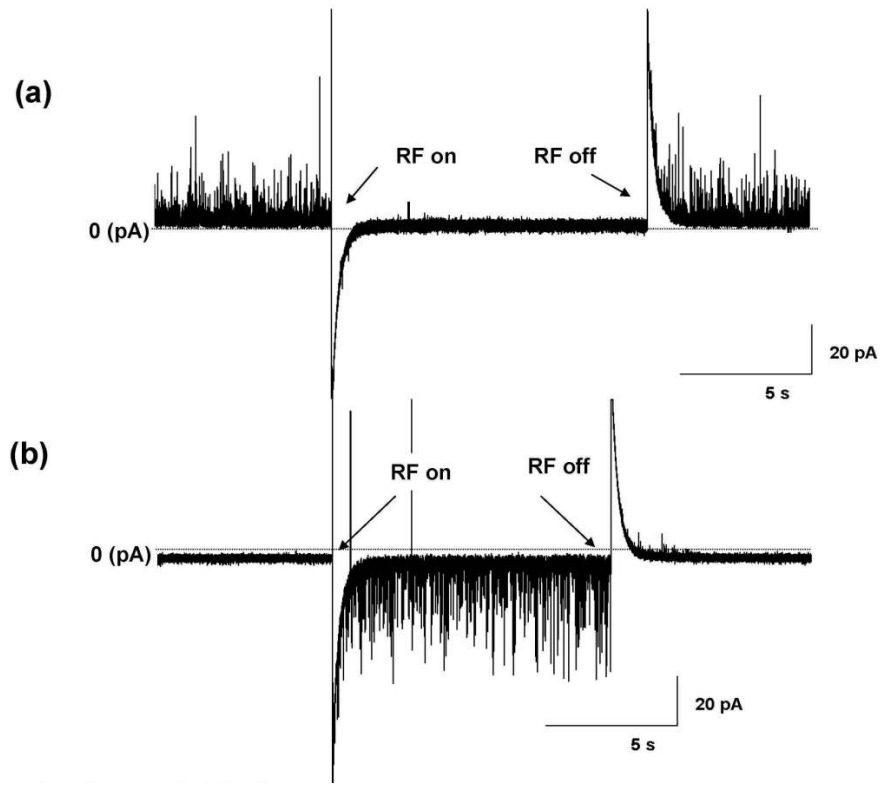

Ramachandran *et al*, Fig. 3/4

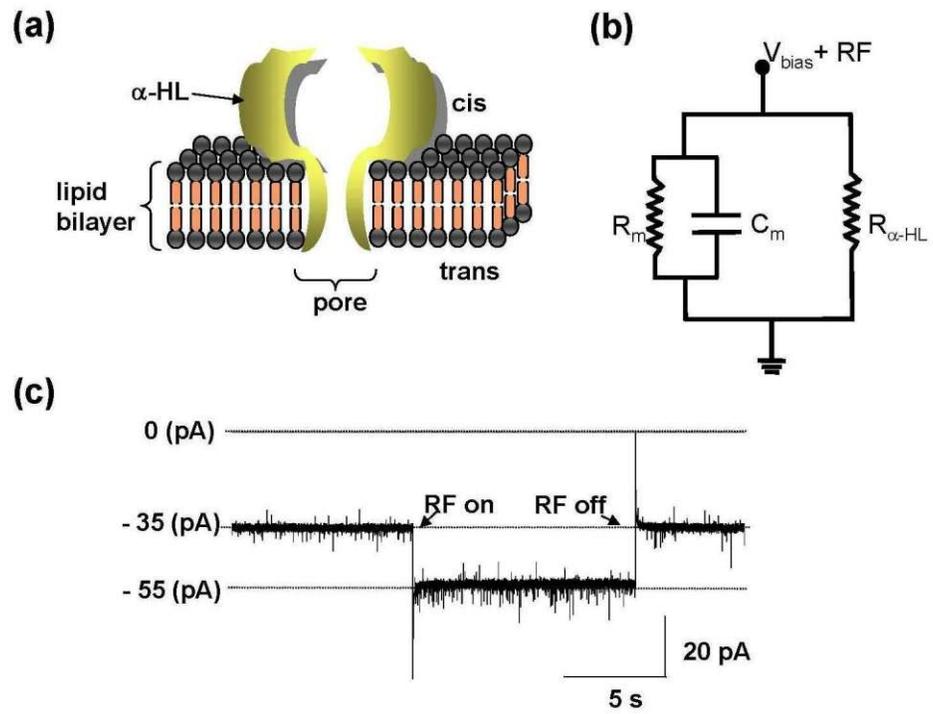

Ramachandran *et al,* Fig. 4/4